# Biexciton in one-dimensional Mott insulators

T. Miyamoto[1], T. Kakizaki[1], T. Terashige[2], D. Hata[1], H. Yamakawa[1], T. Morimoto[1], N. Takamura[1], H. Yada[1], Y. Takahashi[3], T. Hasegawa[4], H. Matsuzaki[2,5], T. Tohyama[6], & H. Okamoto[1,2]

[1]Department of Advanced Materials Science, University of Tokyo, Kashiwa, 277-8561, Japan

[2]AIST-UTokyo Advanced Operando-Measurement Technology Open Innovation Laboratory (OPERANDO-OIL), National Institute of Advanced Industrial Science and Technology (AIST), Chiba 277-8568, Japan

[3]Department of Chemistry, Faculty of Science, Hokkaido University, Sapporo 060-0810, Japan

[4]Department of Applied Physics, University of Tokyo, Tokyo 113-8656, Japan

[5]National Metrology Institute of Japan (NMIJ), National Institute of Advanced Industrial Science and Technology (AIST), Tsukuba, Ibaraki 305-8568, Japan

[6]Department of Applied Physics, Tokyo University of Science, Tokyo 125-8585, Japan

Correspondence and requests for materials should be addressed to T. Miyamoto (email: miyamoto@k.u-tokyo.ac.jp) or H.O. (email: okamotoh@k.u-tokyo.ac.jp)

Mott insulators sometimes show dramatic changes in their electronic states after photoirradiation, as indicated by photoinduced Mott-insulator-to-metal transition. In the photoexcited states of Mott insulators, electron wavefunctions are more delocalized than in the ground state, and long-range Coulomb interactions play important roles in charge dynamics. However, their effects are difficult to discriminate experimentally. Here, we show that in a one-dimensional Mott insulator, bis(ethylenedithio)tetrathiafulvalene-difluorotetracyanoquinodimethane (ET-F$_2$TCNQ), long-range Coulomb interactions stabilize not only excitons, doublon-holon bound states, but also biexcitons. By measuring terahertz-electric-field-induced reflectivity changes, we demonstrate that odd- and even-parity excitons are split off from a doublon-holon continuum. Further, spectral changes of reflectivity induced by a resonant excitation of the odd-parity exciton reveals that an exciton-biexciton transition appears just below the exciton-transition peak. Theoretical simulations show that long-range Coulomb interactions over four sites are necessary to stabilize the biexciton. Such information is indispensable for understanding the non-equilibrium dynamics of photoexcited Mott insulators.

The ultrafast dynamics of correlated electron systems after photoexcitation are now attracting considerable attention. This is based upon recent developments in femtosecond laser technology, which enabled us to detect ultrafast electronic responses to a light pulse in solids[1-4]. Applications of femtosecond pump-probe spectroscopy to correlated electron systems enable us to observe exotic photoinduced phase transitions[5–18] as represented by a photoinduced Mott-insulator-to-metal transition[6,12,16–18] and also to derive detailed information about the interplays between the charge, spin, and lattice degrees of freedom from the transient responses of each degree to a light pulse[5–11,13–15]. Such information can hardly be obtained from the steady-state transport and magnetic measurements. The growing interest in the ultrafast dynamics of correlated electron systems synchronizes to the development of a new field called 'non-equilibrium quantum physics in solids'. In fact, new theoretical approaches have recently been explored to analyse the charge, spin, and lattice dynamics of non-equilibrium states after photoirradiation, as exemplified by the dynamical mean field theory[19] and the time-dependent density-matrix renormalization group method[20,21].

In the non-equilibrium quantum physics of correlated electron systems, the charge dynamics of photoexcited Mott insulators is the most fundamental subject to be studied from both the experimental[6,12,16–18] and theoretical viewpoints[22–28]. Recent studies have focussed on Mott insulator states realized not only in solids such as transition metal compounds[6,16-18] and organic molecular materials[12] but also in ultra-cold atoms on an optical lattice[29,30,31]. In fact, in the ultra-cold atoms, non-equilibrium dynamics can be investigated by tuning the intersite interaction using a Feshbach resonance[29]. Among various Mott insulators, a one-dimensional (1D) Mott insulator with large on-site Coulomb repulsion energy $U$ is particularly important since the charge and spin degrees of freedom are decoupled[32,33], In the system, we can obtain clear information about the effects of Coulomb interactions on the charge dynamics. When the electronic structure and low-energy excitations in a 1D Mott insulator are theoretically analysed, the Hubbard model, which includes $U$ and the transfer energy $t$ as the important parameters, is generally used. In the photoexcited states, on the other hand, electron wavefunctions are more delocalized, and the effects of long-range Coulomb interactions will become important in the charge dynamics. However, it is difficult to evaluate these effects experimentally.

Based upon these backgrounds, in the present study, we investigated the role of long-range Coulomb interactions in photoexcited states by focusing on the excitons and biexcitons in 1D Mott insulators. The long-range Coulomb interactions can stabilize not only the bound state of a doublon and a holon (that is, an exciton[34–36]) but also the bound state of two excitons (that is, a biexciton). The schematics of an exciton and biexciton in a half-filled 1D Mott insulator are shown in Fig. 1a. Assuming that the transfer energy $t$ is equal to zero, a simplified understanding of the stabilities of excitons and biexcitons is possible as follows. Here, we consider the intersite Coulomb repulsive energies for the nearest, second-nearest, and third-nearest two sites, which are denoted by $V_1$, $V_2$, and $V_3$, respectively, as shown in Fig. 1a. In this case, the energy of the lowest exciton is $U - V_1$, and the binding energy of the exciton, that is, the Coulomb attractive energy of a doublon (D) and one holon (H) is $-V_1$. The energy of two isolated excitons far distant from each other is $2(U - V_1)$, while that of a neighbouring two excitons (DHDH) is $(2U - 3V_1 + 2V_2 - V_3)$, thus giving the stabilization energy or binding energy of the biexciton as $(V_1 - 2V_2 + V_3)$. When we assume that the Coulomb repulsion energies are inversely proportional to the distance between two electrons, $V_2 = V_1/2$ and $V_3 = V_1/3$, the binding energy of the biexciton is $V_1/3$ accordingly, and the biexciton is stable as well as the exciton. Since the binding energy of $V_1/3$ originates from $V_3$, an observation of a biexciton can give valuable information about the role of long-range Coulomb interactions in the photoexcited states of 1D Mott insulators. If the long-range Coulomb interaction is important, it will modify the charge dynamics of 1D Mott insulators, e.g., the efficiency of photoinduced Mott-insulator-to-metal transition and the temporal dynamics during the transition.

The studied material is an organic molecular compound, bis(ethylenedithio)tetrathiafulvalene-difluorotetracyanoquinodimethane (ET-F$_2$TCNQ). This compound is a segregated-stacked charge-transfer (CT) compound consisting of ET (donor) and F$_2$TCNQ (acceptor) columns, as shown in Fig. 1b[37]. An electron is transferred from ET to F$_2$TCNQ. F$_2$TCNQ$^-$ molecules are almost isolated, while a finite overlap of wavefunctions with a transfer energy $t$ of ~0.2 eV exists between neighbouring ET$^+$ molecules along the $a$ axis (Fig. 1b). Because of the large $U$ on ET, the ET columns are in a 1D Mott insulator state. Figure 1c shows the polarized reflectivity spectra. A sharp peak polarized parallel to the $a$ axis $(// a)$ is observed at 0.7 eV, which corresponds to the Mott gap transition. Such a sharp structure makes us expect an excitonic nature. From the spectral shape, however, we cannot determine whether this peak is attributed to an exciton or an interband transition sharpened owing to the Van-Hove singularity.

To investigate the energy-level structures of the photoexcited states and stabilities of excitons in a 1D Mott insulator of ET-F$_2$TCNQ, we performed terahertz-pulse-pump optical-reflectivity-probe spectroscopy and measured the electric-field-induced changes in the optical reflectivity spectrum, which include information not only about one-photon allowed states but also originally one-photon forbidden states. From analyses of the results, we clarified that the odd- and even-parity excitons are split off from the doublon-holon continuum. We next applied pump-probe reflection spectroscopy to ET-F$_2$TCNQ in the near-infrared region with a resonant excitation to the lowest exciton, and investigated the possible bound state of two excitons, that is, a biexciton. We observed the signature of an exciton-biexciton transition in the optical reflectivity spectrum, the spectral shape of which was well reproduced by a theoretical simulation taking into account the Coulomb interactions over up to four sites. The results demonstrate the importance of long-range Coulomb interactions in the dynamics of photoexcited excitons in Mott insulators.

**Results**

**Terahertz-pump optical-reflectivity-probe spectroscopy.**

An effective method to evaluate the energy-level structures of excitons is electroreflectance (ER) spectroscopy[38–41], in which reflectivity changes induced by quasistatic electric fields are measured. This enables us to obtain a wide frequency range of the third-order non-linear susceptibility $\chi^{(3)}$ spectrum without special laser systems. However, the ER spectroscopy cannot be applied to low-resistivity materials, in which an application of high electric fields sometimes gives rise to a dielectric breakdown, destroying the sample owing to excess electric current. In most organic molecular compounds with small gap energies, nonlinear electric transport and current-induced electric-switching phenomena indeed occur. This makes it impossible to adopt the ER method. To overcome this problem, in the present study, we apply terahertz-pump optical-probe spectroscopy to ET-F$_2$TCNQ (Fig. 2a). Within a terahertz pulse, an electric current hardly flows owing to the short duration of the electric field (~1 ps)[42]. In addition, the magnitude of the electric field can be increased without sample damages.

In Fig. 2b, we show a typical electric field waveform $E_{\mathrm{THz}}(t_\mathrm{d})$ of the terahertz pulse as a function of the delay time $t_\mathrm{d}$. The peak magnitude of $E_{\mathrm{THz}}(t_\mathrm{d})$ is ~100 kV/cm. The electric fields $E$ of both the terahertz pulse and the optical probe pulse are set to be polarized parallel to the *a* axis ($E$ // *a*). Figure 2c, d shows the time evolutions of reflectivity changes $\Delta R(t_\mathrm{d})/R$ at 0.72 and 0.80 eV, respectively (open blue circles). Large $\Delta R(t_\mathrm{d})/R$ signals reaching $\sim -3\%$ at 0.72 eV and $\sim +2\%$ at 0.80 eV are observed. These time evolutions are almost in agreement with $\pm[E_{\mathrm{THz}}(t_\mathrm{d})]^2$ as shown by the solid red lines in Fig. 2c,d. In fact, the $|\Delta R(t_\mathrm{d} = 0\ \mathrm{ps})/R|$ values at 0.72 eV and 0.80 eV are proportional to $[E_{\mathrm{THz}}(t_\mathrm{d} = 0\ \mathrm{ps})]^2$ (see Supplementary Note 1). These results indicate that $\Delta R/R$ signals are ascribed to the third-order nonlinear-optical response expressed as follows[43]:

$$P^{(3)}(\omega) \propto E_{\mathrm{THz}}(\omega \sim 0) E_{\mathrm{THz}}(\omega \sim 0) E(\omega), \qquad (1)$$

where $P^{(3)}(\omega)$ and $E(\omega)$ are the third-order nonlinear polarization and the electric field of the probe light, respectively. Such a reflectivity modulation by a terahertz electric field was previously reported in [Ni(chxn)$_2$Br]Br$_2$ (chxn=cyclohexanediamine)[42].

To obtain detailed information about the energy level structure, we measured the probe-energy dependence of $\Delta R/R$ at $t_\mathrm{d} = 0$ ps (open circles in Fig. 2e). By applying the Kramers–Kronig (KK) transformation to the $R$ and $\Delta R(t_\mathrm{d} = 0\ \mathrm{ps})/R$ spectra, we obtained the $\varepsilon_2$ and $\Delta \varepsilon_2$ spectra, as shown by the solid black line in Fig. 2f and solid blue line in Fig. 2g, respectively. Details of the analyses are reported in Supplementary Note 2. The $\varepsilon_2$ spectrum has a sharp peak at 0.7 eV, and the $\Delta \varepsilon_2$ spectrum has a plus-minus-plus structure around the sharp peak.

First, we analyse the $\varepsilon_2$ spectrum with the following Lorentzian-type dielectric function:

$$\varepsilon_2(\omega) = \frac{Ne^2}{\hbar}\langle 0|x|1\rangle^2 \mathrm{Im}\left(\frac{1}{\omega_1 - \omega - i\gamma_1} + \frac{1}{\omega_1 + \omega + i\gamma_1}\right). \quad (2)$$

Here, $|0\rangle$ and $|1\rangle$ show the ground state and the one-photon-allowed odd-parity state, respectively, and $\langle 0|x|1\rangle$ is the transition dipole moment between them. $\omega_1$ and $\gamma_1$ are the energy and damping constants of the odd-parity state, respectively. $N$ denotes the density of the ET molecules, $\varepsilon_0$ is the permittivity of the vacuum, $e$ is the elementary charge, and $\hbar$ is the reduced Planck constant. As shown in Fig. 2f, the experimental $\varepsilon_2$ spectrum (the solid black line) is almost reproduced by Eq. (2), as shown by the red line. The used parameter values are listed in Table I.

We next analysed the $\Delta\varepsilon_2$ spectrum showing a plus-minus-plus structure (the solid blue line in Fig. 2g). To analyse this spectrum, we assumed that the frequency of the terahertz electric field, $\omega_{\mathrm{THz}}$, is 0. This assumption is justified under the condition that $\hbar\omega_{\mathrm{THz}}$ (~4 meV) is sufficiently lower than an energy difference between any of two energy levels of excited states[38]. ET-F$_2$TCNQ meets this condition, as will be shown later. Using this assumption, we calculate $\mathrm{Im}\chi^{(3)}$ from $\Delta\varepsilon_2$ with the following equation:

$$\Delta\varepsilon_2(\omega) = 3\mathrm{Im}\chi^{(3)}(-\omega; 0, 0, \omega)E_{\mathrm{THz}}(0)^2. \quad (3)$$

The maximum of $|\mathrm{Im}\chi^{(3)}|$ was evaluated to be $1 \times 10^{-7}$ esu.

The previous ER spectroscopy of 1D Mott insulators of transition metal compounds revealed that a plus-minus-plus structure in $\mathrm{Im}\chi^{(3)}$ spectra can be interpreted by a three-level model in which the one-photon forbidden state with even parity ($|2\rangle$) is assumed in addition to the ground state $|0\rangle$ and the odd-parity state $|1\rangle$[38–41]. In ET-F$_2$TCNQ, small negative signals appear above ~0.85 eV, as shown in Fig. 2g, in addition to the plus-minus-plus structure. Such a negative signal can be explained by considering the second-lowest odd-parity state ($|3\rangle$)[41]. In a four-level model consisting of $|0\rangle - |3\rangle$, the $\chi^{(3)}$ spectrum is represented by the following equation[43]:

$$\chi^{(3)}(-\omega_\sigma; \omega_i, \omega_j, \omega_k)$$

$$= \frac{Ne^4}{6\varepsilon_0 \hbar^3} \wp \sum_{abc} \left[ \frac{\langle 0|x|a\rangle\langle a|x|b\rangle\langle b|x|c\rangle\langle c|x|0\rangle}{(\omega_a - \omega_\sigma - i\gamma_a)(\omega_b - \omega_j - \omega_k - i\gamma_b)(\omega_c - \omega_k - i\gamma_c)} \right.$$

$$+ \frac{\langle 0|x|a\rangle\langle a|x|b\rangle\langle b|x|c\rangle\langle c|x|0\rangle}{(\omega_a + \omega_i + i\gamma_a)(\omega_b - \omega_j - \omega_k - i\gamma_b)(\omega_c - \omega_k - i\gamma_c)}$$

$$+ \frac{\langle 0|x|a\rangle\langle a|x|b\rangle\langle b|x|c\rangle\langle c|x|0\rangle}{(\omega_a + \omega_i + i\gamma_a)(\omega_b + \omega_i + \omega_j + i\gamma_b)(\omega_c - \omega_k - i\gamma_c)}$$

$$\left. + \frac{\langle 0|x|a\rangle\langle a|x|b\rangle\langle b|x|c\rangle\langle c|x|0\rangle}{(\omega_a + \omega_i + i\gamma_a)(\omega_b + \omega_i + \omega_j + i\gamma_b)(\omega_c + \omega_\sigma + i\gamma_c)} \right]. \quad (4)$$

$\langle l|x|m\rangle$ shows the transition dipole moment between states $|l\rangle$ and $|m\rangle$. $\omega_l$ and $\gamma_l$ are the frequency and the damping constant for the state $|l\rangle$, respectively. $\wp$ is the permutation of $(\omega_i, \omega_j, \omega_k)$. $|a\rangle$ and $|c\rangle$ show odd-parity states, and $|b\rangle$ shows an even-parity state. A schematic of the four-level model is shown in Fig. 2h. Among the combinations of $(a, b, c)$, the terms with $b = 0$ are much smaller than the others. Thus, they can be excluded, and $(a, b, c) = (1, 2, 1), (1, 2, 3), (3, 2, 1), (3, 2, 3)$ should be considered. Using Eqs. (3) and (4), the $\text{Im}\chi^{(3)}$ spectrum is almost reproduced, as shown by the red line in Fig. 2g. The energies of three excited states ($\hbar\omega_1 = 0.694$ eV, $\hbar\omega_2 = 0.720$ eV, and $\hbar\omega_3 = 0.850$ eV) are indicated by triangles in the same figure. The obtained parameters are also listed in Table I.

The splitting between $|1\rangle$ and $|2\rangle$, $\hbar(\omega_2 - \omega_1)$, was small (26 meV), indicating that the two states are nearly degenerate. In addition, $\langle 1|x|2\rangle$, which is the most important parameter dominating the magnitude of $\chi^{(3)}$, was very large at ~18 Å. In ET-F$_2$TCNQ, the ratio $\langle 1|x|2\rangle/\langle 0|x|1\rangle$ is equal to 13. The enhancement of $\langle 1|x|2\rangle$ is attributable to the fact that the wave functions of the odd- and even-parity states are similar to each other except for their phases, as schematically shown in Fig. 2h, and the spatial overlap of these wave functions becomes very large. These features are characteristic of 1D Mott insulators[38,39]. The observation of a higher odd-parity state $|3\rangle$ demonstrates that the lower two states $|1\rangle$ and $|2\rangle$ are excitionc states.

To obtain evidence of the excitonic effect from the transport property, we measured the excitation profile of photoconductivity (PC), which is shown by open circles in Fig. 2f. The PC signals are very low at $\hbar\omega_1$ and $\hbar\omega_2$, suggesting that $|1\rangle$ and $|2\rangle$ are excitonic states. With an increase in the photon energy, the PC increases and saturates at around $\hbar\omega_3 = 0.850$ eV. In the energy region of the doublon-holon continuum, a number of both odd and even states exist continuously. Thus, field-induced reflectivity changes $\Delta R/R$ originating from their mixings cancel each other, and $\Delta R/R$ signals appear only at the lower edge of the continuum state[41]. Therefore, it is reasonable to consider that the continuum state starts at around $\hbar\omega_3 = 0.850$ eV, and that $\hbar(\omega_3 - \omega_1) = 0.156$ eV and $\hbar(\omega_3 - \omega_2) = 0.130$ eV are crude measures of the binding energies of the lowest $|1\rangle$ and second-lowest excitonic states $|2\rangle$, respectively.

**Optical-pump optical-reflectivity-probe spectroscopy.**

To observe a biexciton, we next performed optical-pump optical-reflectivity-probe spectroscopy (Fig. 3a) by resonant excitation of the lowest odd-parity exciton ($\hbar\omega_1 = 0.694$ eV). Electric fields of the pump pulses were polarized parallel to the *a* axis as well as the probe pulses with *E // a*.

Figure 3b shows the time evolutions of $\Delta R/R$ for three typical probe energies: 0.58 eV, 0.694 eV, and 0.96 eV. We set the excitation fluence $I_{ex}$ to be 5.1 µJ/cm², which corresponds to the excitation photon density $x_{ph}$ of 0.0015 photons per ET molecule. This excitation photon density is low enough to detect the transition of an isolated exciton to a biexciton state. Excitation photon density dependence of exciton-biexciton transition is detailed in Supplementary Note 3. The reflectivity at 0.694 eV corresponding to the exciton peak decreases just after the photoirradiation, and most of the change recovers within 0.15 ps. Such an ultrafast change and recovery of $\Delta R/R$ is not observed by the higher-energy excitation at 1.55 eV as previously reported (see Supplementary Note 4)[44]. Therefore, this response is attributable to the coherent response, which is observed in the resonant excitation to an exciton in semiconductors[45]. That can be interpreted as a kind of third order nonlinear responses such as an optical Stark effect and a stimulated emission[45]. These responses may become important when a probe pulse is incident to a sample within a phase coherence time of an electronic polarization induced by a pump pulse. Besides the ultrafast component due to the coherent response, $\Delta R/R$ at 0.694 eV should also include the bleaching signal owing to the real excitations of excitons as a component with a finite decay time. $\Delta R/R$ at 0.96 eV, which is higher than the exciton peak, also seems to partly include the ultrafast coherent response as well as the bleaching signal. By contrast, $\Delta R/R$ at 0.58 eV below the exciton peak increases after the photoirradiation and is accompanied by an oscillatory component, which will be discussed in detail later.

The probe energy dependences of $\Delta R$ at $t_d = 0.3$ ps, $1.0$ ps, and $5.0$ ps are shown in Fig. 3c, together with the original $R$ spectrum. For these delay times, the coherent responses mentioned above almost disappear, and the $\Delta R$ spectra can reflect the effects of real exciton excitations. $\Delta R$ is negative around the original exciton peak (~0.7 eV) owing to the bleaching of the exciton transition. In the lower energy region below ~0.64 eV, $\Delta R$ becomes positive, as seen in the time evolution of $\Delta R/R$ at 0.58 eV in Fig. 3b. To obtain the information about photoexcited states, we calculated the photoinduced change of $\varepsilon_2$ ($\Delta\varepsilon_2$) by the KK transformation of the $\Delta R/R$ spectrum at $t_\mathrm{d} = 0.3$ ps. The details of the analyses are reported in Supplementary Note 2. The obtained $\Delta\varepsilon_2$ spectrum is shown in Fig. 3d together with the original $\varepsilon_2$ spectrum. In addition to a negative $\Delta\varepsilon_2$ around the lowest exciton peak, a positive $\Delta\varepsilon_2$ peak is clearly observed at 0.630 eV. Such a photoinduced absorption is not observed in the case of higher-energy excitation at 1.55 eV[12]. This indicates that the observed photoinduced absorption is characteristic of the lowest-energy exciton. A possible origin of this peak is the transition of the lowest-energy exciton to a biexciton.

The energy difference $\Delta E$ (~60 meV) between the original exciton peak in $\varepsilon_2$ (0.694 eV) and the photoinduced absorption peak in $\Delta\varepsilon_2$ (0.630 eV) corresponds to the binding energy of the biexciton, as shown in Fig. 3e. As mentioned in the introductory part, a simplified model with $t = 0$ shows that the stabilization energy of the biexciton is one third of the intersite Coulomb interaction $V_1$, $-V_1/3$. The binding energy of the lowest odd-parity exciton (~160 meV) is expected to be almost equal to $V_1$. Thus, the biexciton binding energy is estimated to be $V_1/3 \sim 53$ meV, which is in accord with $\Delta E \sim 60$ meV. This supports the validity of our interpretation that the photoinduced absorption is attributed to the biexciton.

**Simulation of exciton-biexciton transition**

To investigate the biexciton formation more strictly, we theoretically calculate the imaginary part of the dielectric constant $\varepsilon_2$ in the ground state and in the presence of the lowest exciton using an extended Hubbard model, as follows:

$$H = -t \sum_{i,\sigma} \left( e^{iA(t)} C_{i,\sigma}^{\dagger} C_{i+1,\sigma} + \text{H.c.} \right) + U \sum_{i} \left( n_{i,\uparrow} - \frac{1}{2} \right)\left( n_{i,\downarrow} - \frac{1}{2} \right)$$

$$+ V_1 \sum_{i} (n_i - 1)(n_{i+1} - 1) + V_2 \sum_{i} (n_i - 1)(n_{i+2} - 1)$$

$$+ V_3 \sum_{i} (n_i - 1)(n_{i+3} - 1) \qquad (5)$$

Here, $V_j$ is the Coulomb interaction between two electrons distant for $j$ sites, as mentioned above. We assume again that $V_j$ is inversely proportional to a doublon-holon distance as $V_1 : V_2 : V_3 = 1 : \frac{1}{2} : \frac{1}{3}$. In Fig. 3f, we show the $\varepsilon_2$ spectrum in the ground state, which was calculated by the Lanczos method with a system size (site number) of $L$ = 14. The parameter values in the system are set to be $t = 0.14$ eV, $U = 1.4$ eV, and $V_1 = 0.6$ eV to reproduce the peak energy of the $\varepsilon_2$ spectrum for the odd-parity exciton.

Next, we calculated the $\varepsilon_2$ spectrum after the resonant excitation to the odd-parity exciton[42]. The temporal shape of the pump pulse is assumed to be Gaussian, as follows:

$$A_{\text{pump}}(t) = A_0 e^{-t_d^2/2\tau^2} \cos(\omega_0 t_d) \qquad (6)$$

$\omega_0$, $A_0$, and $\tau$ are the frequency, amplitude, and temporal width of the pump pulse, respectively. We calculated $\Delta\varepsilon_2$ with the parameter values of $\omega_0 = 0.694$ eV and $\tau = 78$ fs, which correspond to a full-width half-maximum of 130 fs. The result for $t_d = 0.3$ ns is shown by the solid blue line in Fig. 3f, which reproduces the important feature of the experimental $\Delta\varepsilon_2$ spectrum, that is, the presence of the induced absorption just below the original absorption peak at 0.630 eV attributable to the exciton-biexciton transition. The exciton-biexciton transition owing to long-range Coulomb interactions was also theoretically predicted in 2D Mott insulators[47]. From an increase in the energy in the system, we evaluated the photocarrier density $\delta$ to be 0.003, which ensures a weak excitation condition. These theoretical calculations ascertain that the biexciton as well as the exciton are stable in 1D Mott insulators.

**Exciton relaxation observed as coherent oscillations**

In this subsection, we discuss the results of the time evolutions and the probe-energy dependence of the oscillatory component $\Delta R_{\text{OSC}}/R$ observed in the photoinduced reflectivity changes $\Delta R/R$. In Fig. 4a, we show a typical time characteristic of $\Delta R_{\text{OSC}}/R$ by open circles, which is extracted from the time evolution of $\Delta R/R$ at 0.58 eV (see Supplementary Note 5). In this experiment, the excitation photon density $x_{\text{ph}}$ is set at 0.01 photon per ET molecule. This oscillation is almost reproduced by the convolution of a damped oscillator expressed below and the Gaussian profile corresponding to the time resolution (150 fs) as shown by the solid red line in Fig. 4a.

$$\frac{\Delta R_{\text{OSC}}}{R} = A_{\text{OSC}} \exp\left(-\frac{t_\text{d}}{\tau}\right) \cos(\omega_{\text{OSC}} t + \phi) \tag{7}$$

The oscillation frequency $\omega_{\text{OSC}}$ is 82 cm$^{-1}$, and the relaxation time $\tau$ is 2.0 ps. The value of $\phi$ ($= 14°$) is small, suggesting that the generation mechanism of the oscillation is the displacive excitation of the coherent phonon[48]. The Fourier power spectrum of the experimental time characteristic of $\Delta R_{\text{OSC}}/R$ and the fitting curve are shown in Fig. 4b by open circles and the solid red line, respectively, which agree with each other.

We performed similar analyses of the coherent oscillations in $\Delta R/R$ signals at various probe energies and plotted the magnitude of the fitting functions ($A_{\text{OSC}}$) in Fig.

4c (red circles) together with the original $R$ spectrum (black line). The data show a clear peak at 0.64 eV, which corresponds well to the peak (0.630 eV) of $\Delta\varepsilon_2$ assigned to the exciton-to-biexciton transition shown in Fig. 3d. This suggests that the energy and/or the intensity of the exciton-to-biexciton transition is modulated at a frequency of 82 cm$^{-1}$, which is observed as the oscillatory structure of the reflectivity changes. The origin of this oscillation is discussed in the next section.

**Discussion**

First, we comment on the stabilization mechanism of the biexciton in 1D Mott insulators. In the simulation with an extended Hubbard model, we investigated several parameter sets. When we consider only the intersite Coulomb interactions $V_1$ and $V_2$, no peak is observed just below the lowest exciton transition in $\Delta\varepsilon_2$, even if their magnitudes are enhanced. This demonstrates that the long-range Coulomb attractive interaction characterized by $V_3$ plays a significant role in the stabilization of the biexciton. This is consistent with the simplified picture of the energy gain of biexciton formation in Fig. 1a.

Next, we discuss the origin of the coherent oscillation. As seen in the spectrum of the magnitude of the oscillatory components in Fig. 4c, the oscillation is observed around the exciton-biexciton transition at 0.630 eV (Fig. 3d). In addition, the frequency of the oscillation, 82 cm$^{-1}$, is a typical frequency of a lattice mode in organic molecular compounds. It is, therefore, reasonable to consider that the oscillation is ascribed to molecular displacements in the lattice relaxation process of the lowest-energy exciton generated by the resonant excitation at 0.694 eV. A possible origin is the molecular displacement associated with the molecular dimerization stabilizing the exciton as illustrated in Fig. 4d, which corresponds to the phonon mode with the wavenumber $k =$ $\pi/a_0$. Here, $a_0$ is the lattice constant along the $a$ axis. Such dimeric molecular displacements and the subsequent coherent oscillation are considered to be produced over several sites around the exciton and change the intersite Coulomb attractive interactions $V_1$ and $V_3$ that stabilizes the biexciton. Note that in a region at a distance from the sites where excitons are produced by the pump pulse, no oscillations are generated, so that coherent oscillations are hardly detected around the peak energy (0.7 eV) of the original exciton transition. Thus, the oscillation modulates the energy and intensity of the exciton-biexciton transition, which results in an oscillatory signal $\Delta R_{\mathrm{OSC}}/R$ on $\Delta R/R$ only around the exciton-biexciton transition.

The time evolution of the induced absorption due to the exciton-biexciton transition (the positive $\Delta R/R$ signal at 0.58 eV in Fig. 3b) and that of the bleaching signal due to the real excitations of excitons (the negative $\Delta R/R$ signals at 0.694 eV and 0.96 eV in Fig. 3b) should reflect the decay dynamics of excitons. To evaluate the exciton decay dynamics, we performed the fitting analyses on the time evolutions of $\Delta R/R$ at 0.58 eV, 0.694 eV, and 0.96 eV. The details of the analyses are reported in the Supplementary Note 6.

The results showed that the time evolutions of $\Delta R/R$ at 0.694 eV and 0.96 eV consist of the fast component with the decay time $\tau_{\mathrm{fast}}$ of 0.39 ps and the slow component with the decay time $\tau_{\mathrm{slow}}$ of 8.6 ps. The value of $\tau_{\mathrm{fast}}$ is consistent with those reported in the previous experimental studies[44,49] and in the theoretical calculations[50]. This ultrafast decay of excitons may be attributed to the emission of intramolecular vibrations. Their frequencies range from 500 to 1500 cm$^{-1}$, so that an exciton with the energy of ~0.7 eV (~5600 cm$^{-1}$) can decay via the emission of several high-frequency phonons. The slow decay component with the decay time $\tau_{\mathrm{slow}}$ of 8.6 ps can be related to the lattice relaxation of excitons. When the exciton is relaxed by the dimeric molecular displacements, the exciton is better stabilized and the decay time becomes longer possibly up to several picoseconds. This decay time is still very short compared to a decay time of excitons in inorganic semiconductors, which is on the order of nanoseconds. The analysis also revealed that the time evolution of $\Delta R/R$ at 0.58 eV also includes the slow decay component with the decay time $\tau_{\text{slow}}$ of 8.3 ps. The fast decay component with the decay time $\tau_{\text{fast}}$ of 0.39 ps might also be included in the $\Delta R/R$ signal at 0.58 eV, however, it cannot be discriminated owing to the presence of the large negative signal coming from the coherent response. We note that the relaxation of excitons due to dimeric molecular displacements strengthens the intensity and decreases the energy of the exciton-biexciton transition via an increase in the intersite Coulomb attractive interactions $V_1$ and $V_3$ by the decrease in the distance between the nearest and third-nearest two molecules, respectively.

In summary, we successfully measured the spectra of the ultrafast reflectivity changes by a terahertz electric field and by the resonant excitation of the lowest exciton in a 1D Mott insulator of an organic molecular compound, ET-F$_2$TCNQ. By analysing the spectra of reflectivity changes induced by the terahertz electric field, we revealed the energy-level structures of the exciton and continuum states, and evaluated the binding energy of the lowest-energy exciton to be about 160 meV. In addition, from the spectrum of reflectivity changes by the resonant optical excitation to excitons, we demonstrated that the biexciton is stable owing to long-range Coulomb interactions and that its binding energy is about 60 meV, which is almost equal to one third of the exciton binding energy as predicted in a case with a strong electron correlation limit. Such information about biexcitons as well as excitons is significantly important for the understanding of non-equilibrium dynamics in photoexcited 1D Mott insulators.

**Materials and Methods**

**Sample preparations.**

Single crystals of ET-F$_2$TCNQ were grown by the method previously reported[37].

**Pump-probe reflectivity measurements.**

In the terahertz-pulse-pump optical-reflectivity-probe experiments, the output of Ti:Al$_2$O$_3$ regenerative amplifier (pulse width: 130 fs, photon energy: 1.58 eV, repetition frequency: 1 kHz) were divided into two beams. One was used for the generation of terahertz-pump pulses through the optical rectification in a LiNbO$_3$ crystal by the pulse-front-tilting method[51,52]. The other was introduced to an optical parametric amplifier (OPA) to obtain probe lights from 0.48 to 1.08 eV. A temporal waveform of terahertz electric field $[E_{\text{THz}}(t)]$ was measured by the electro-optical sampling with a 1 mm-thick (110) oriented ZnTe crystal[42]. Magnitude of the terahertz electric field was changed by two wire-grid polarizers and a Si plate inserted in the terahertz-pulse path. The time origin for the terahertz-pump experiments is set at the maximum of the terahertz electric field amplitude. Time difference between pump and probe pulses $t_{\text{d}}$ was controlled by a changing the path length of the probe pulse. Polarizations of all pulses are parallel to the 1D stacking-axis *a*.

In the optical-pump experiments, the output of Ti:Al$_2$O$_3$ regenerative amplifier (pulse width: 110 fs, photon energy: 1.55 eV, repetition frequency: 1 kHz) was also divided to two beams. These two beams were respectively introduced to two OPAs, and the pump and probe pulses were generated. The excitation photon density $x_{\text{ph}}$ was evaluated from the average within the penetration depth of the pump pulse ($l_{\text{ex}}$) using the equation $x_{\text{ph}} = (1 - R_{\text{ex}})(1 - 1/e)I_{\text{ex}}/l_{\text{ex}}$, where $R_{\text{ex}}$ and $I_{\text{ex}}$ are the reflectivity and intensity per unit area of the pump pulse, respectively. Time difference between pump and probe pulses $t_{\text{d}}$ was controlled by changing the path length of the pump pulse. Polarizations of all pulses are also parallel to the 1D stacking-axis $a$.

All the experiments were performed at 294 K.

**Photoconductivity measurements**

The photoconductivity (PC) measurement was performed by the method previously reported[40] at 92 K to avoid thermal excitations of carriers. We ascertained that the photocurrents are proportional to the intensity of the excitation light. The excitation spectrum of PC was obtained by using the formula, $I_{\text{PC}} \propto A_{\text{PC}}/[I_{\text{p}}(1 - R_{\text{p}})]$. Here, $A_{\text{PC}}$, $I_{\text{P}}$, and $R_{\text{p}}$ are the photo-current signal, the photon density of the excitation light per unit area, and the reflection loss.

**Data availability.**

The data that support the findings of this study are available from the corresponding author on request.

**41,** 2326- 2338 (1990).

33. Eskes, H. & Oleś, A. M. Two Hubbard Bands: Weight Transfer in Optical and One-Particle Spectra. *Phys. Rev. Lett.* **73,** 1279-1282 (1994).

34. Stephan, W. & Penc, K. Dynamical density-density correlations in one-dimensional Mott insulators. *Phys. Rev. B* **54,** R17269-R17272 (1996).

35. Essler, F. H. L., Gebhard, F. & Jeckelmann, E. Excitons in one-dimensional Mott insulators. *Phys. Rev. B* **64,** 125119 (2001).

36. Jeckelmann, E. Optical excitations in a one-dimensional Mott insulator. *Phys. Rev. B* **67,** 075106 (2003).

37. Hasegawa, T., Kagoshima, S., Mochida, T., Sugiura, S. & Iwasa, Y. Electronic states and anti-ferromagnetic order in mixed-stack charge-transfer compound (BEDT-TTF)(F$_2$TCNQ). *Solid State Commun.* **103,** 489-493 (1997).

38. Kishida, H. et al. Gigantic optical nonlinearity in one-dimensional Mott–Hubbard insulators. *Nature* **405,** 929-932 (2000).

39. Mizuno, Y., Tsutsui, K., Tohyama, T. & Maekawa, S. Nonlinear optical response and spin-charge separation in one-dimensional Mott insulators. *Phys. Rev. B* **62,** R4769-R4773 (2000).

40. Ono, M. et al. Linear and nonlinear optical properties of one-dimensional Mott insulators consisting of Ni-halogen chain and CuO-chain compounds. *Phys. Rev. B* **70,** 085101 (2004).

41. Ono, M., Kishida, H. & Okamoto, H. Direct Observation of Excitons and a Continuum of One-Dimensional Mott Insulators: A Reflection-Type Third-Harmonic-Generation Study of Ni-Halogen Chain Compounds. *Phys. Rev. Lett.* **95,** 087401 (2005).
23

amplitudes exceeding 1 MV/cm generated by optical rectification in $LiNbO_3$. *Appl. Phys. Lett.* **98,** 091106 (2011).

**Acknowledgements**

This work was supported in part by Grants-in-Aid for Scientific Research from the Japan Society for the Promotion of Science (JSPS) (Project Numbers: JP25247049, JP18K13476, JP18H01166) and by CREST (Grant Number: JPMJCR1661), Japan Science and Technology Agency. T. Morimoto, H. Yamakawa, and T. Terashige were supported by the Japan Society for Promotion of Science through Program for Leading Graduate Schools (MERIT). T. Morimoto and H. Yamakawa were supported by a fellowship of the Japan Society for the Promotion of Science.

**Author contributions**

T. Miyamoto, T. Terashige, D.H., H. Yamakawa, T. Morimoto, N.T., H. Yada, and N.K. constructed the terahertz-pump optical-reflectivity-probe systems. T. Miyamoto, T. Terashige and N.T. performed terahertz-pump optical-reflectivity-probe measurements. T. Miyamoto, T.K. and H.M. carried out optical-pump optical-reflectivity-probe measurements. Y.T. and T.H. provided single crystals of ET-$F_2$TCNQ. T. Tohyama performed theoretical calculations. H.O. coordinated the study. All of the authors discussed the results and contributed to writing the paper.

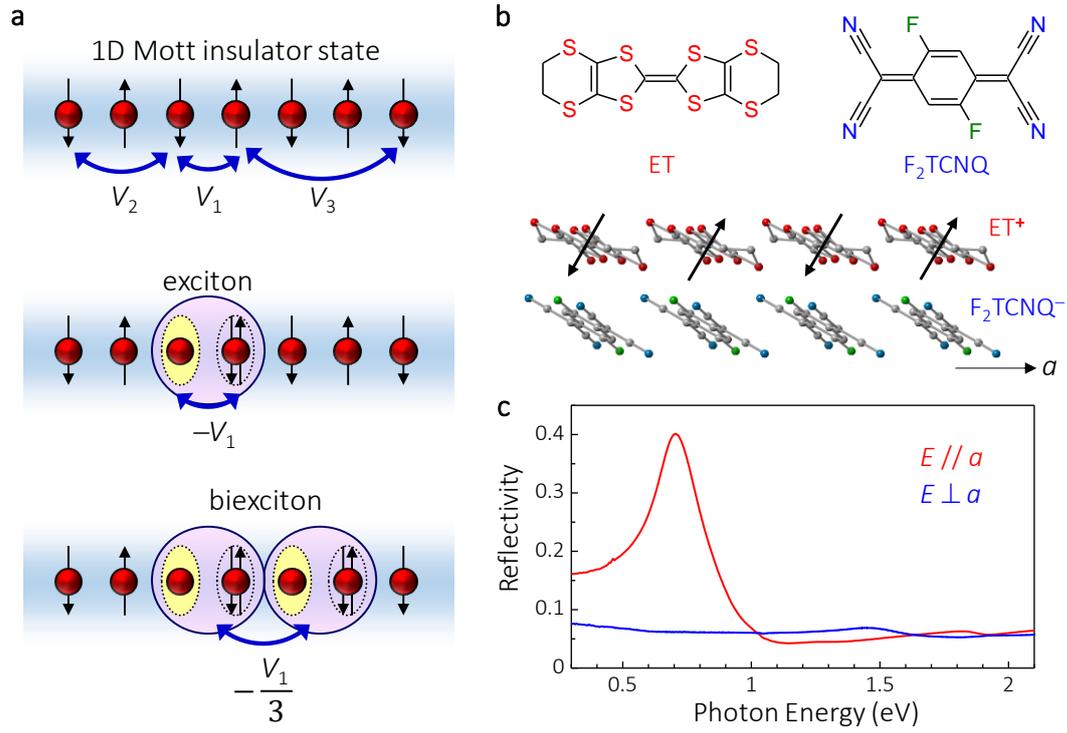

**Fig. 1** Photoexcited states in a 1D Mott insulator of ET-F$_2$TCNQ. **a** Schematics of a ground state, exciton, and biexciton in a half-filled 1D Mott insulator. The binding energy of biexciton is $(V_1 - 2V_2 + V_3)$, which is equal to $V_1/3$ in the case that the Coulomb repulsion energies are inversely proportional to the distance between two electrons, $V_2 = V_1/2$ and $V_3 = V_1/3$. **b** Molecular structures and 1D molecular stacks along the *a* axis of ET-F$_2$TCNQ. **c** Polarized reflectivity spectra of ET-F$_2$TCNQ with *E*//*a* and *E*⊥*a*.

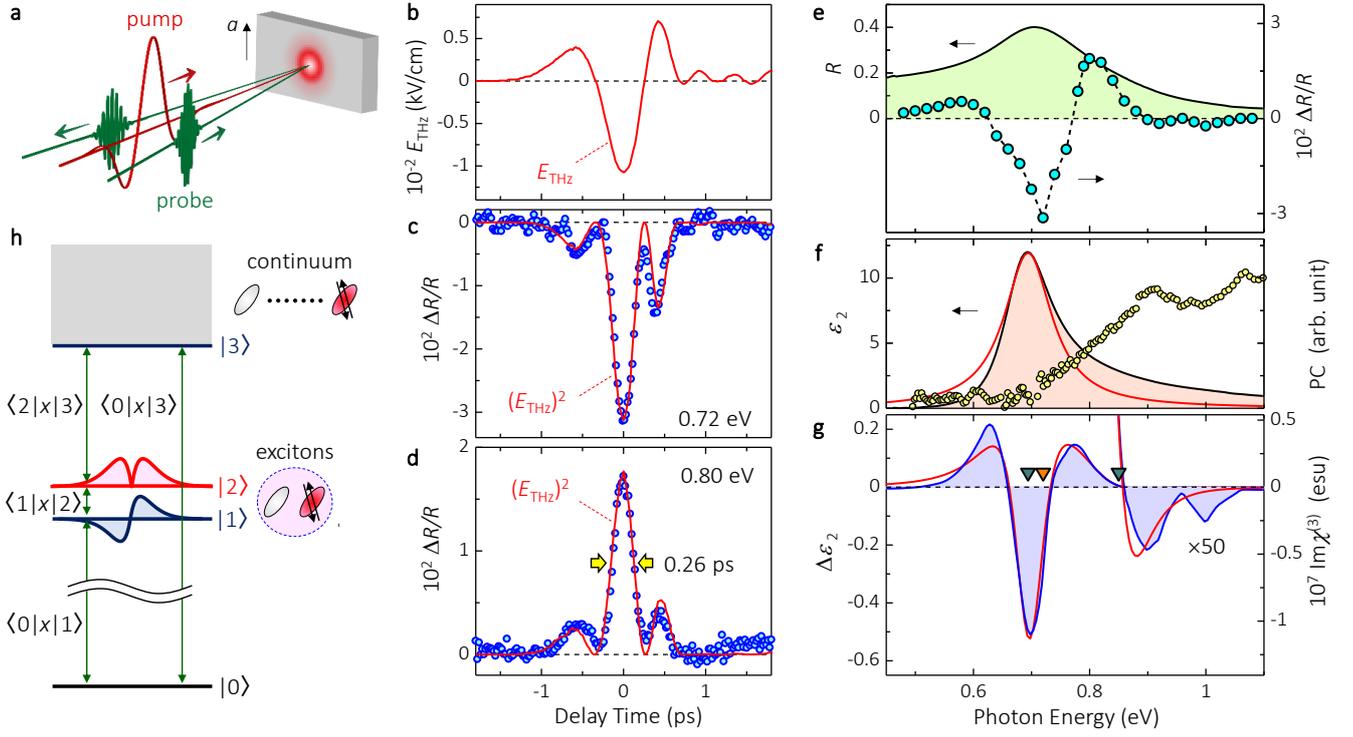

**Fig. 2** Terahertz-pulse-pump optical-reflectivity-probe spectroscopy. **a** A schematic of the experimental setup. **b** A typical waveform of $E_{\mathrm{THz}}$. **c,d** Time evolutions of $\Delta R/R$ at (**c**) 0.72 eV and (**d**) 0.80 eV (blue open circles). The red solid lines show the normalized value of $|E_{\mathrm{THz}}|^2$. **e** A reflectivity spectrum polarized along the *a* axis (solid line) and $\Delta R/R$ spectrum at 0 ps (open circles). **f** $\varepsilon_2$ spectrum (black solid line), a fitting curve to $\varepsilon_2$ spectrum (red solid line), and photoconductivity (PC) spectrum (open circles). **g** $\Delta\varepsilon_2$ ($\mathrm{Im}\chi^{(3)}$) spectrum (blue solid line) and a fitting curve to $\mathrm{Im}\chi^{(3)}$ spectrum (red solid line). Open triangles show the energy levels of the excited states. **h** A schematic of the four-level model.

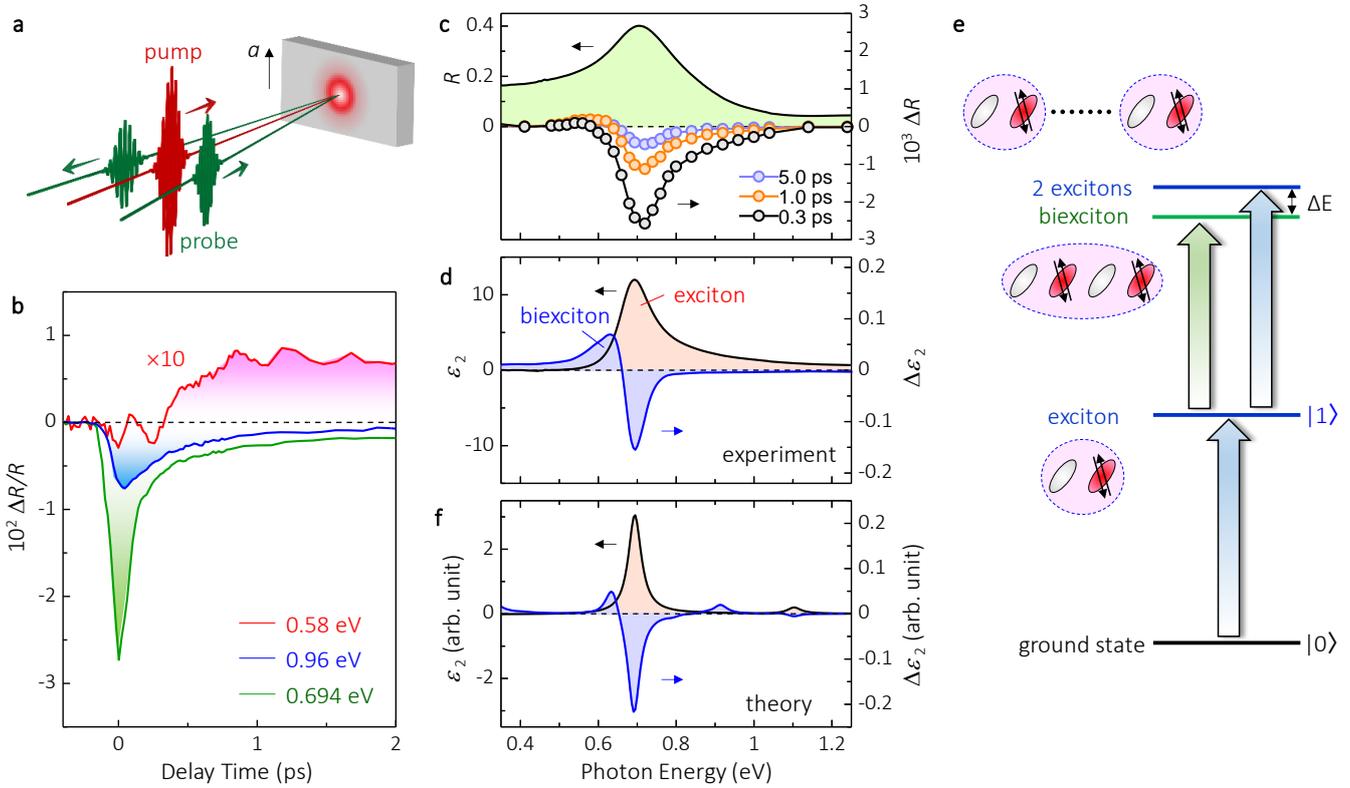

**Fig. 3** Optical-pump optical-reflectivity-probe spectroscopy. **a** A schematic of the experimental setup. **b** Time evolutions of the reflectivity changes at 0.58 eV (red solid line), 0.694 eV (green solid line) and 0.96 eV (blue solid line). **c** $\Delta R$ spectrum at 0.3 ps (black open circles), 1.0 ps (red open circles) and 5.0 ps (blue open circles). The black solid line shows a polarized reflectivity spectrum with $E//a$. **d** $\varepsilon_2$ spectrum (black solid line) and $\Delta\varepsilon_2$ spectrum (blue solid line) from the experimental results. **e** A schematic of the excitations of exciton, biexciton and 2 excitons. **f** $\varepsilon_2$ spectrum (black solid line) and $\Delta\varepsilon_2$ spectrum (blue solid line) obtained by the theoretical calculations.

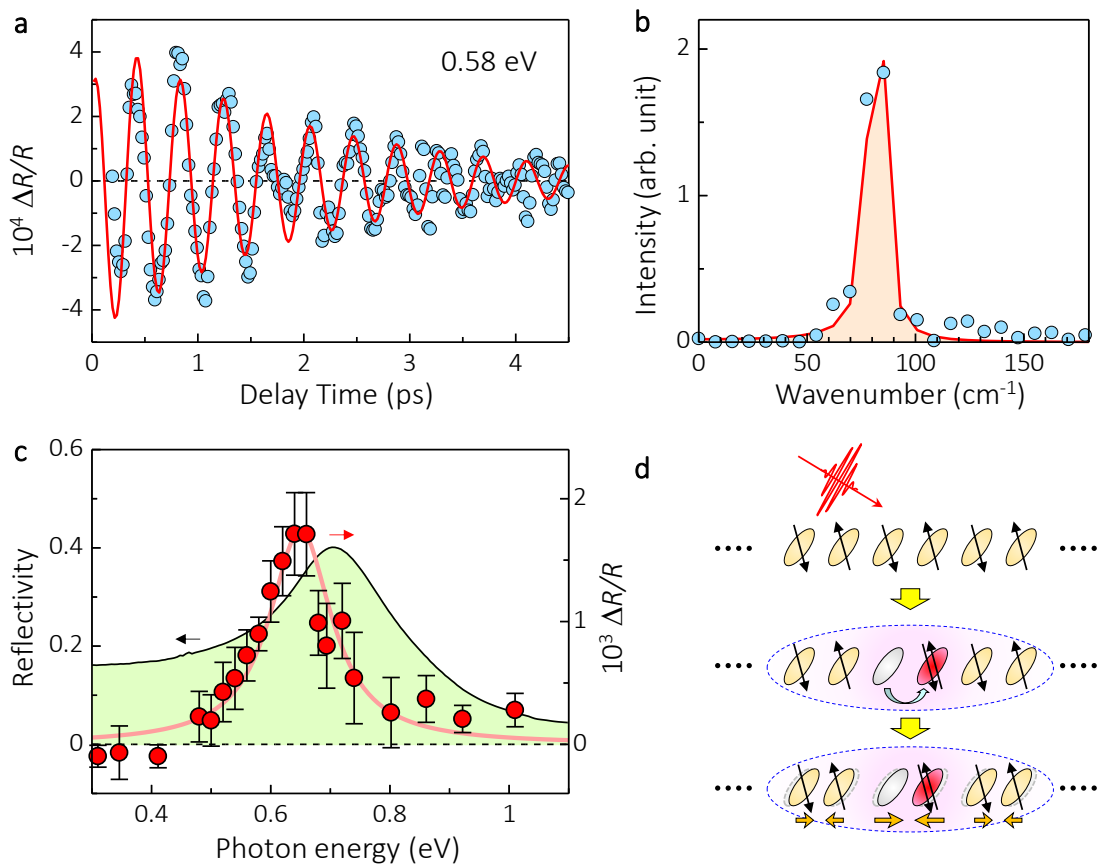

**Fig. 4** Coherent oscillations after photo-excitation. **a,b** Open circles show a time evolution of the oscillatory component at 0.58 eV (**a**) and its Fourier transformation (**b**). The red solid lines show fitting curves. **c** Magnitude of the fitting functions ($A_{\mathrm{OSC}}$) (open circles) together with the original $R$ spectrum with $E//a$ (the black line). **d** A schematic of the coherent oscillations after photo-excitation.

Table I. Parameters evaluated from the fitting analysis.

| $\langle 0|x|1\rangle$ | $\langle 1|x|2\rangle$ | $\langle 2|x|3\rangle\langle 3|x|0\rangle$ | $\hbar\omega_1$ | $\hbar\omega_2$ | $\hbar\omega_3$ | $\hbar\gamma_1$ | $\hbar\gamma_2$ | $\hbar\gamma_3$ |
|---|---|---|---|---|---|---|---|---|
| 1.4 Å | 18 Å | −7.1 Å² | 0.694 eV | 0.720 eV | 0.850 eV | 0.050 eV | 0.146 eV | 0.081 eV |

**Supplementary Information for**

**Biexciton in one-dimensional Mott insulators**

T. Miyamoto[1], T. Kakizaki[1], T. Terashige[2], D. Hata[1], H. Yamakawa[1], T. Morimoto[1], N. Takamura[1], H. Yada[1], Y. Takahashi[3], T. Hasegawa[4], H. Matsuzaki[2,5], T. Tohyama[6], & H. Okamoto[1,2]

[1]*Department of Advanced Materials Science, University of Tokyo, Kashiwa, 277-8561, Japan*

[2]*AIST-UTokyo Advanced Operando-Measurement Technology Open Innovation Laboratory (OPERANDO-OIL), National Institute of Advanced Industrial Science and Technology (AIST), Chiba 277-8568, Japan*

[3]*Department of Chemistry, Faculty of Science, Hokkaido University, Sapporo 060-0810, Japan*

[4]*Department of Applied Physics, University of Tokyo, Tokyo 113-8656, Japan*

[5]*National Metrology Institute of Japan (NMIJ), National Institute of Advanced Industrial Science and Technology (AIST), Tsukuba, Ibaraki 305-8568, Japan*

[6]*Department of Applied Physics, Tokyo University of Science, Tokyo 125-8585, Japan*

**Supplementary Note 1. Terahertz electric-field dependence of reflectivity changes**

To clarify whether the observed reflectivity changes by terahertz electric fields originate from the third-order optical nonlinearity, we measured $\Delta R(t_\text{d} = 0 \text{ ps})/R$ for various amplitudes of the terahertz electric fields $E_\text{THz}(0)$. In Supplementary Figure 1, we show the $E_\text{THz}(0)$ dependence of $|\Delta R(t_\text{d} = 0 \text{ ps})/R|$ at probe energies of 0.72 eV and 0.80 eV. Both data items are proportional to the square of $E_\text{THz}(0)$ up to ~110 kV/cm. This demonstrates that the observed $\Delta R/R$ signals can be attributed to the third-order optical nonlinearity.

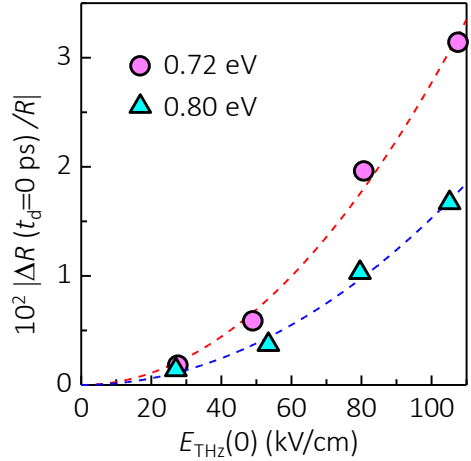

**Supplementary Figure 1.** Terahertz-electric-field $E_\text{THz}(0)$ dependence of the reflectivity changes $|\Delta R(t_\text{d} = 0 \text{ ps})/R|$ at 0.72 eV and 0.80 eV (open circles and triangles, respectively). Broken lines show the quadratic dependences on $E_\text{THz}(0)$.

**Supplementary Note 2. Kramers–Kronig transformation of reflectivity-change spectra**

To perform the Kramers–Kronig (KK) transformations of the $\Delta R/R$ spectra, we extrapolated the $\Delta R/R$ spectra since the measurement range was limited. In the $\Delta R/R$ spectra obtained by terahertz-pulse-pump optical-reflectivity-probe spectroscopy, the measurement range was 0.48 eV to 1.08 eV. Below (above) the lower (higher) energy bound of the measured range, we linearly extrapolated the data using the lowest (highest) two measurement points and set $\Delta R/R$ to be zero after $\Delta R/R$ reached zero. We also made the liner interpolation in the $\Delta R/R$ spectra experimentally obtained from 0.48 eV to 1.08 eV since the number of measurement points was insufficient to perform the KK transformation.

In the $\Delta R/R$ spectra obtained by optical-pump optical-reflectivity-probe spectroscopy, the measurement range was 0.082 eV to 2.17 eV. Below the lower energy bound of the measurement range, we used an $\Delta R/R$ value of 0.082 eV. Above the higher energy bound, we linearly extrapolated the data using the highest two measurement points and set $\Delta R/R$ to be zero after $\Delta R/R$ reached zero. We also linearly interpolated the $\Delta R/R$ spectra from 0.082 eV to 2.17 eV.

**Supplementary Note 3. Excitation photon density dependence of exciton-biexciton transition**

To clarify how the dynamics of biexcitons depends on the density of excitons generated by the pump pulse, we measured the excitation photon density $x_{\text{ph}}$ dependence of reflectivity changes $\Delta R/R$ at 0.58 eV, which corresponds to the peak energy of the exciton-biexciton transition. Supplementary Figure 2 shows $\Delta R/R$ at 0.58 eV ($t_{\text{d}}$= 1 ps) as a function of $x_{\text{ph}}$. The magnitudes of $\Delta R/R$ is proportional to $x_{\text{ph}}$ below $x_{\text{ph}} = 0.015$ photons (ph)/ET and then saturated. The observed saturation of the signals for $x_{\text{ph}} > 0.015$ ph/ET suggests that under the presence of a high density of excitons or equivalently doublon-holon pairs, the subsequent probe pulse is difficult to form a stable biexciton. We consider that a high density of excitons would destabilize a biexciton by modifying the long-range Coulomb interactions possibly through screening or many-body effects. Taking these facts into account, we investigated the exciton-biexciton transition in the low-fluence excitation condition with $x_{\text{ph}} = 0.0015$ ph/ET.

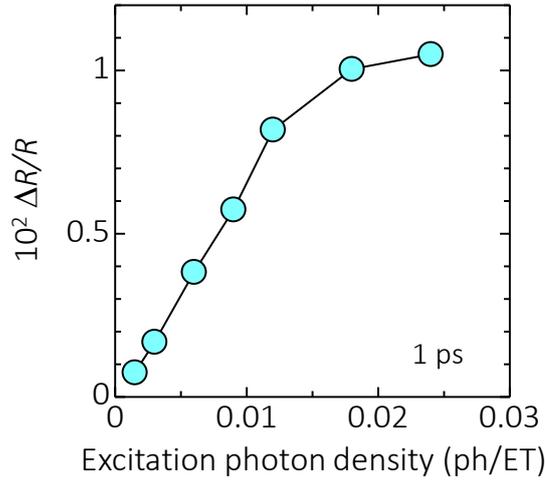

**Supplementary Figure 2.** Reflectivity changes $\Delta R/R$ at 0.58 eV ($t_d = 1$ ps) as a function of $x_{\text{ph}}$.

**Supplementary Note 4. Excitation-photon-energy dependence of time evolutions of reflectivity change**

To distinguish between the coherent response and the bleaching signal owing to the real excitation of excitons, we compared the time evolution of $\Delta R/R$ by the resonant excitation to the lowest odd-parity exciton and that by the higher-energy excitation with 1.55 eV as previously reported[1]. These are indicated by the green and red solid lines, respectively, in Supplementary Figure 3. In the case of the 1.55-eV excitation, no ultrafast component can be observed around the time origin in contrast to the case of the resonant exciton excitation. This indicates that the ultrafast component of $\Delta R/R$ does not originate from the real excitation of excitons nor carriers but the coherent response. On the other hand, both time evolutions include a slower decay component with a relaxation time of several ps, so this can be ascribed to the real excitation of excitons or carriers.

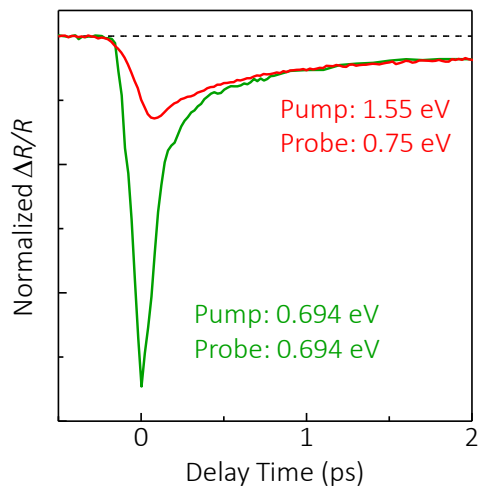

**Supplementary Figure 3.** Time evolutions of the reflectivity change at 0.694 eV with the resonant excitation to the lowest odd-parity exciton (green solid line) and at 0.75 eV with the higher energy excitation with 1.55 eV[1] (red solid line).

**Supplementary Note 5. Analyses of coherent oscillations on photoinduced reflectivity changes**

To analyse the coherent oscillations on the time evolutions of $\Delta R/R$ obtained by the optical-pump optical-reflectivity-probe measurements, we derived the oscillatory components by applying a high-pass Fourier filter with a cut-off frequency of 1.6 THz to the $\Delta R/R$ data. To exclude the contribution of the coherent response around the time origin, we used the data only for $t_\text{d} > 0.2$ ps. We next performed fitting analyses of the oscillatory components thus obtained using eq. (7) as described in the main text. In these analyses, we used the common values of decay time $\tau = 2.0$ ps and oscillation frequency $\omega_\text{osc} = 82 \text{ cm}^{-1}$.

**Supplementary Note 6. Evaluation of decay dynamics of excitons**

To evaluate the decay dynamics of the exciton, we performed the fitting analysis of the time evolutions of $\Delta R/R$ at 0.694 eV and 0.96 eV by using the following function convolved with the Gaussian profile corresponding to the time resolution (150 fs).

$$\Delta R/R\,(t) = A\delta(t) + A_{\text{fast}}\exp\left(-\frac{t}{\tau_{\text{fast}}}\right) + A_{\text{slow}}\exp\left(-\frac{t}{\tau_{\text{slow}}}\right) \qquad (S1)$$

The first term represents the coherent response and its magnitude is denoted by $A$. $\delta(t)$ is the delta function. The second (third) term shows the fast (slow) decay component of excitons, the magnitude and decay time of which are denoted by $A_{\text{fast}}$ ($A_{\text{slow}}$) and $\tau_{\text{fast}}$ ($\tau_{\text{slow}}$), respectively.

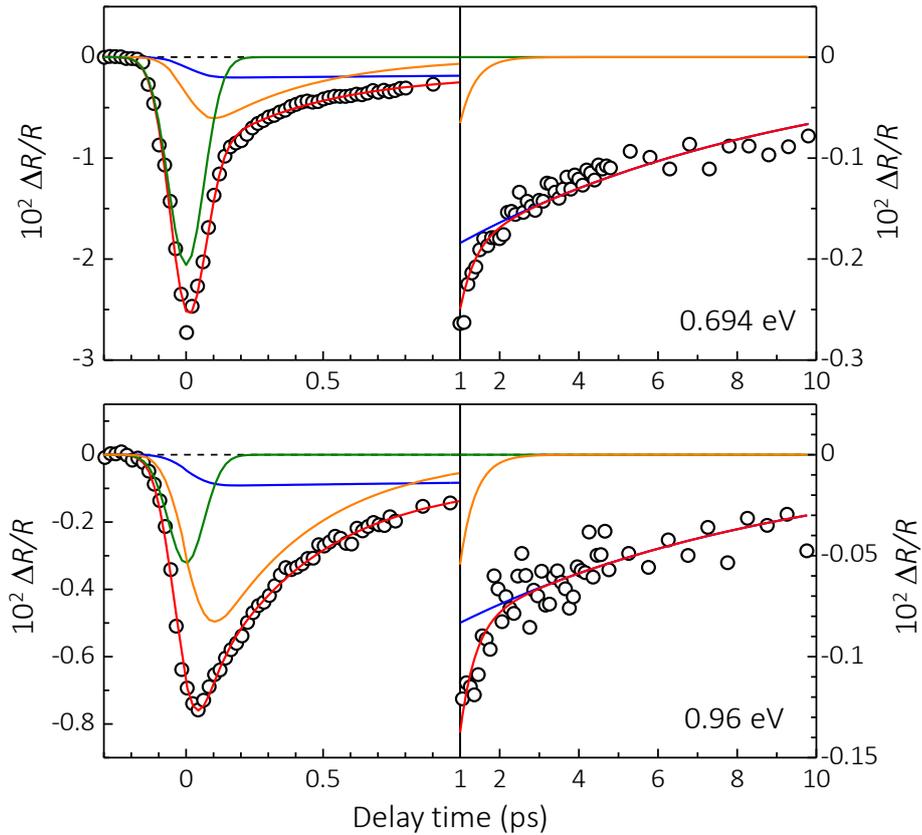

**Supplementary Figure 4.** Time evolutions (circles) and fitting curves (red lines) of reflectivity changes at 0.694 and 0.96 eV. The green, orange, and blue lines represent the coherent response, the fast decay component, and the slow decay component, respectively.

The time evolutions of $\Delta R/R$ at 0.694 eV and 0.96 eV, and the fitting curves are shown in Supplementary Fig. 4. Both time evolutions are well reproduced by the fitting curves as shown by the red solid lines with the common values of $\tau_{fast}$=0.39 ps and $\tau_{slow}$=8.6 ps. The time evolution of the coherent response, the fast decay component, and the slow decay component are also shown by the green, orange, and blue lines, respectively, in the same figure.

The positive $\Delta R/R$ signal at 0.58 eV due to the exciton-biexciton transition also gives the information about the decay dynamics of exciton. As mentioned in the main text, the $\Delta R/R$ signal at 0.58 eV will include the contribution of the negative $\Delta R/R$ component due to the coherent response around the time origin, so that we analysed only the data for $t_d > 1$ ps using the following formula including only the third term in Eq. (S1).

$$\Delta R/R\,(t) = A_{slow}\exp\left(-\frac{t}{\tau_{slow}}\right) \qquad (S2)$$

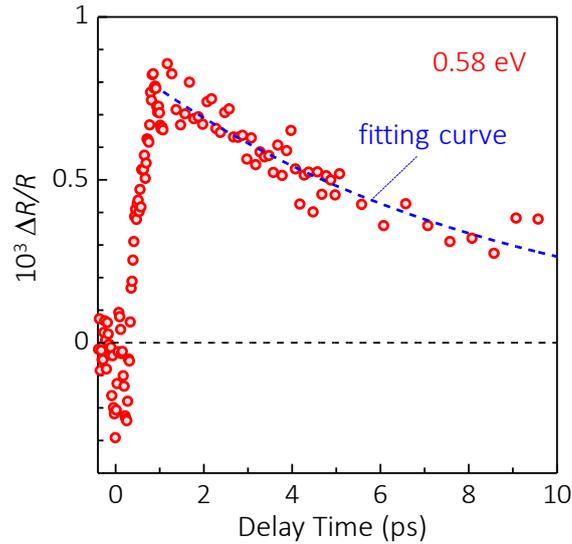

**Supplementary Figure 5.** A time evolution of the reflectivity change at 0.58 eV (red open circles) and a fitting curve (blue broken line).

In Supplementary Fig. 5, we show the time evolution of $\Delta R/R$ at 0.58 eV and the fitting curve. The data is well reproduced by Eq. (S2) as shown by the blue broken line. The decay time $\tau_{\text{slow}}$ is 8.3 ps, which is almost the same as the values (8.6 ps) obtained by the fitting analyses of the data at 0.694 eV and 0.96 eV. It is reasonable since the time evolution of the exciton-biexciton transition reflects the decay characteristic of excitons.